\documentclass[aps,pra,reprint,amsmath,amssymb,superscriptaddress,floatfix,showpacs]{revtex4-1}

\usepackage{graphicx}
\usepackage{hyperref}
\usepackage[english]{babel}

\makeatletter
\let\l@English\l@english
\makeatother
\newcommand{\kpc}{\chi_{pc}}
\newcommand{\kcp}{\chi_{cp}}
\newcommand{\kpp}{\chi_{pp}}
\newcommand{\kcc}{\chi_{cc}}

\begin{document}

\title{Role of the phase-matching condition in non-degenerate four-wave mixing 
in hot vapors for the generation of squeezed states of light}

\author{M. T. Turnbull}

\author{P. G. Petrov}

\author{C. S. Embrey}

\affiliation{Midland Ultracold Atom Research Centre, School of Physics and Astronomy,
University of Birmingham, Edgbaston, Birmingham B15 2TT, UK}

\author{A. M. Marino}
\affiliation{Homer L. Dodge Department of Physics and Astronomy, The
University of Oklahoma, 440 W. Brooks St., Norman, Oklahoma 73019, USA}

\author{V. Boyer}

\affiliation{Midland Ultracold Atom Research Centre, School of Physics and Astronomy,
University of Birmingham, Edgbaston, Birmingham B15 2TT, UK}

\date{\today}

\begin{abstract}

    Non-degenerate forward four-wave mixing in hot atomic vapors has been
    shown to produce strong quantum correlations between twin beams of light
    [McCormick et al, \textit{Opt.\ Lett.} \textbf{32}, 178 (2007)], in a
    configuration which minimizes losses by absorption. In this paper, we look
    at the role of the phase-matching condition in the trade-off that occurs
    between the efficiency of the nonlinear process and the absorption of the
    twin beams. To this effect, we develop a semi-classical
    model by deriving the atomic susceptibilities in the relevant
    double-lambda configuration and by solving the classical propagation of
    the twin-beam fields for parameters close to those found in typical
    experiments. These theoretical results are confirmed by a simple
    experimental study of the nonlinear gain experienced by the twin beams as
    a function of the phase mismatch. The model shows that the amount of phase
    mismatch is key to the realization of the physical conditions in which the
    absorption of the twin beams is minimized while the cross-coupling between
    the twin beams is maintained at the level required for the generation of
    strong quantum correlations. The optimum is reached when the four-wave
    mixing process is not fully phase matched.

\end{abstract}


\maketitle

\section{Introduction}

Continuous-variable entanglement can be generated deterministically with a
phase-insensitive optical amplifier. For a gain larger than 1, the system
produces a two-mode squeezed state where the signal and the idler (here
referred to as probe and conjugate respectively) display EPR-type
entanglement~\cite{reid_demonstration_1989}. Such an amplifier can be realized
using a nonlinear optical process such as parametric
down-conversion~\cite{heidmann_observation_1987,ou_realization_1992} or
four-wave mixing (4WM)~\cite{silberhorn_generation_2001}. In real physical
systems, the presence of absorption reduces the amount of quantum correlations
which can be generated, and although 4WM in atomic vapors can lead to large
gains, resonant atomic processes are also responsible for losses, which limit the
amount of observable squeezing.  Recently, a configuration in 4WM was found
which reduces absorption
\cite{hemmer_efficient_1995,lukin_resonant_1998,*lukin_resonant_2000}, and
generates large degrees of
squeezing~\cite{mccormick_strong_2008,glorieux_quantum_2011,
jasperse_relative_2011,qin_compact_2012}.  

There have been a number of reasons put forward to explain this success. The
main one is the nature of the nonlinearly, which is based on coherence effects
between the hyperfine electronic ground states rather than on the saturation
of a transition of a two-level
atom~\cite{reid_squeezing_1985,lukin_resonant_1998, *lukin_resonant_2000}.
Indeed, avoiding a large atomic population in the excited state is key to the
reduction of the noise associated with spontaneous emission.  More
specifically, it was pointed out that the D1 line of alkali atoms is
particularly amenable to the establishment of a ground state
coherence~\cite{pooser_quantum_2009}. We show here that the production of
squeezing is also due in great part to a judicious choice of the parameters
that most greatly influence the phase-matching condition of the nonlinear
process, specifically the relative frequencies of the beams and the angle
between the beams. Maybe surprisingly, the highest levels of squeezing are
achieved when the system is not fully phase matched.

The paper is divided as follows. In section~\ref{theory}, we theoretically
study 4WM in an atomic vapor in a double-lambda configuration where both pumps
are detuned from the atomic resonance. From the atomic susceptibilities we
evaluate the impact of the phase-matching condition on the gain and the
absorption in the forward-4WM configuration. In section~\ref{experiment}, we
report on a systematic experimental study of the phase-matching condition
which confirms the findings of section~\ref{theory}. In section~\ref{doppler},
we extend our model to take into account the Doppler effect due to the thermal
motion of the atoms.  Finally in section~\ref{quantum} we discuss the impact
of our theoretical and experimental findings on the possibility of generating
strong two-mode squeezing with 4WM in a hot vapor. From the model we deduce
the best parameters in terms of beam geometry and beam detunings, and compare
them to the recent squeezing experiments of
Refs.~\onlinecite{mccormick_strong_2008,glorieux_quantum_2011,
jasperse_relative_2011,qin_compact_2012}.

\section{Theoretical model of non-degenerate four-wave mixing}
\label{theory}

When describing nonlinear media in the presence of off-resonant fields, it is
common to separate the response of the system into a linear part and a
nonlinear part~\cite{boyd_nonlinear_2008}. The linear contribution leads to an
index of refraction which modifies the linear dispersion relation for each of
the individual light fields in the medium. The nonlinear part acts as a
perturbative source term in the propagation equations.  It enables energy
transfer between light fields for those configurations where the
phase-matching condition is fulfilled, that is to say when the total
wave-vector of the waves giving up energy equals the total wave-vector of the
waves receiving the energy. The relevant wave-vectors are those in the medium.
They are equal to the wave-vectors in vacuum times their corresponding indices
of refraction.

In resonant media, such as atomic vapors excited close to atomic transitions,
the medium is strongly perturbed by the presence of the light. For instance,
optical pumping by a strong pump beam can affect the atomic populations in the
hyperfine levels, leading to a strong change in the index of refraction seen
by a weaker beam. In this case, the usual expansion separating the linear and
the nonlinear responses may not be appropriate. Instead we consider an
expansion of the nonlinear polarization to first order in the electric fields
of the weak beams and to all orders in the electric fields of the pump
beams~\cite{lukin_resonant_1998,*lukin_resonant_2000}.

\begin{figure}[htb]
    \begin{center}
	\includegraphics[width=.6\linewidth]{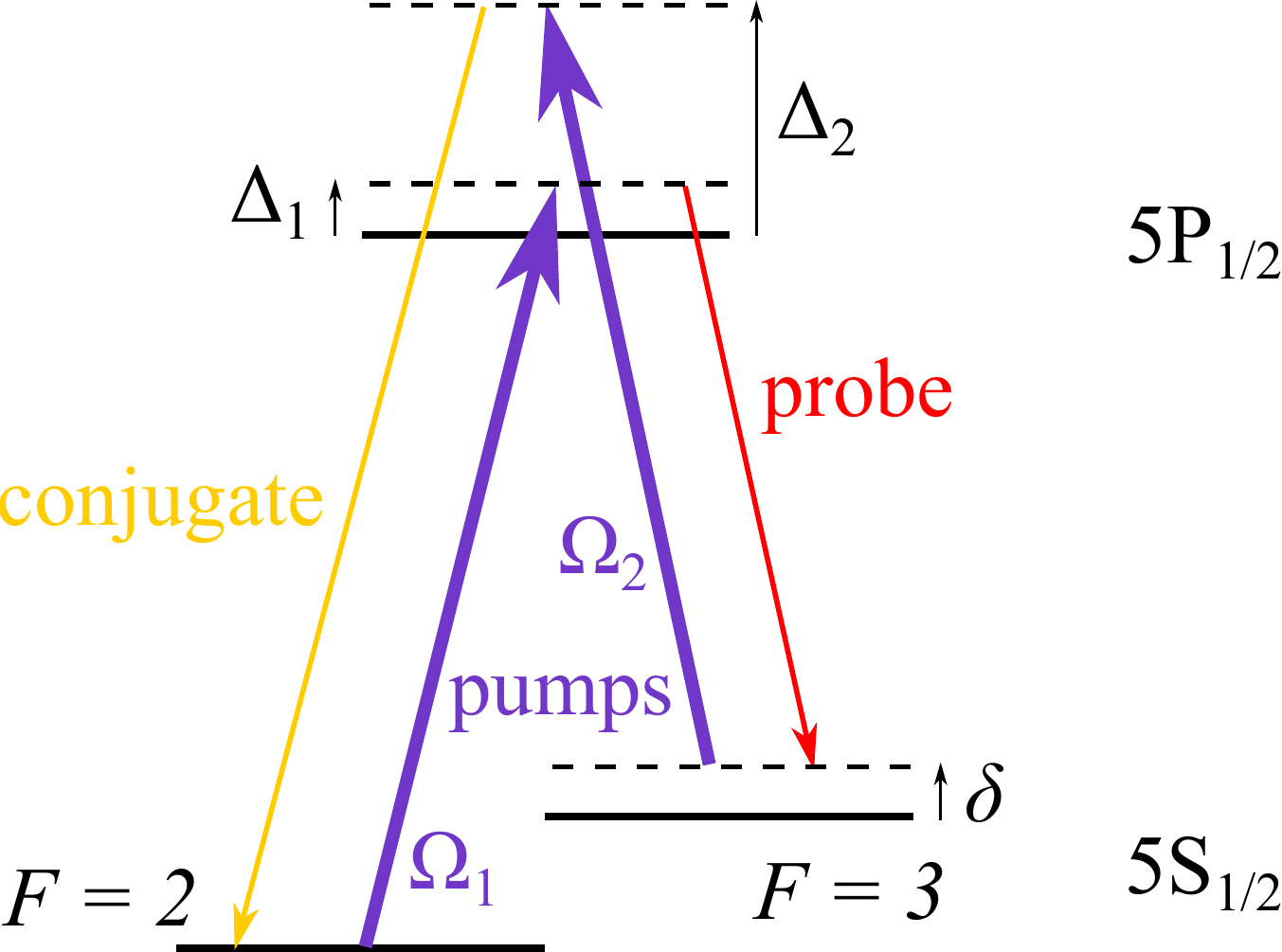}
    \end{center}
    \caption{Double-lambda scheme on the D1 line of $^{85}$Rb. The hyperfine
    splitting of the excited state is not resolved due to Doppler
    broadening. The lambda transitions,
    detuned by $\Delta_1$ and $\Delta_2$, are driven by pump fields
    with resonant Rabi frequencies $\Omega_1$ and 
    $\Omega_2$, respectively. 
    Note that the pumps can actually be a single laser beam.
    }     
    \label{fig:lambda}
\end{figure}

We consider the double-lambda configuration, shown in Fig.~\ref{fig:lambda},
which was used to demonstrate intensity-difference squeezing and quadrature
entanglement in a vapor of
$^{85}$Rb~\cite{mccormick_strong_2007,boyer_entangled_2008}. A non-degenerate
4WM parametric process drives an atom from one of the hyperfine ground states
to the other hyperfine ground state and back to the initial state. In the
process, two pump photons are converted into two twin photons, called probe
and conjugate, with wave-vectors $\mathbf{k}_p$ and $\mathbf{k}_c$ in vacuum,
and frequencies $\omega_p$ and $\omega_c$. The non-linearity originates in a
strong coupling between the probe and the conjugate fields mediated by the
coherence of the electronic ground states~\cite{reid_squeezing_1985}.
Following the usual experimental configuration, we further assume that the two
pump photons come from a single field, of wave-vector $\mathbf{k}_0$ in vacuum
and frequency $\omega_0$. We denote $\delta$ the two-photon detuning of the
pump and the probe: $\delta = \omega_0 - \omega_p - \omega_\mathrm{HF}$ where
$\omega_\mathrm{HF}$ is the hyperfine splitting of the ground state. The
natural linewidth of the excited state is $\gamma = 2\pi \times 6$ MHz. In the
rest of the paper, we vary $\delta$ by changing the frequency of the probe.
The 4WM resonance occurs roughly at $\delta = 0$ (more on this below).  

For simplicity, we assume that the coupling strengths of the pump to both
transitions are equal, corresponding to a resonant Rabi frequency $\Omega =
\Omega_1 = \Omega_2$.  Following Ref.~\onlinecite{lukin_resonant_1998,*lukin_resonant_2000}, we calculate
the atomic susceptibilities at the probe and conjugate frequencies in the
limit of weak probe and conjugate fields. The detailed calculation is
developed in the appendix. The susceptibilities are derived by
calculating the steady-state value of the density matrix of a 4-level system
interacting with the four fields of the double lambda. The atomic polarization
at the frequencies of the probe and the conjugate is proportional to the
average oscillating atomic electric dipole at those frequencies and therefore
to the off-diagonal components of the density matrix corresponding to these
transitions. To first order in the probe and conjugate fields, the atomic
polarization is described by two direct susceptibilities, $\kpp$ and $\kcc$,
and two cross-susceptibilities, $\kpc$ and $\kcp = \kpc^*$, given by
Eqs.~(\ref{dbllambdachi1}--\ref{dbllambdachi4}). The cross-susceptibilities are
responsible for the 4WM.
%
%\begin{eqnarray} P(\mathbf{r}, \omega_{p}) & = &
%\epsilon_{0}\chi_{pp}(\omega_{p})\mathcal{E}_{p}+
%\epsilon_{0}\chi_{pc}(\omega_{p})e^{i\Delta \mathbf{k} \cdot \mathbf{r}
%}\mathcal{E}_{c}^{*} \label{probepol}\\ P(\mathbf{r}, \omega_{c}) & = &
%\epsilon_{0}\chi_{cc}(\omega_{c})\mathcal{E}_{c}+\epsilon_{0}\chi_{cp}(\omega_{c})e^{i\Delta
%\mathbf{k} \cdot \mathbf{r}}\mathcal{E}_{p}^{*}, \label{conjpol}
%\end{eqnarray}
%
 
The propagation equations for the
slowly varying envelopes of the probe and conjugate fields $\mathcal{E}_{p}$
and $\mathcal{E}_{c}$, using the polarization expressions
(\ref{probepol}) and (\ref{conjpol}), are given in steady state by:
\begin{eqnarray} 
    \frac{\partial}{\partial z}\mathcal{E}_{p} & = & \frac{i
    k_p}{2\varepsilon_0}P(\omega_p)e^{-i \mathbf{k}_p \cdot 
    \mathbf{r}}\\ 
    \frac{\partial}{\partial z}\mathcal{E}_{c} & = &
    \frac{i k_c}{2\varepsilon_0}P(\omega_c)e^{-i \mathbf{k}_c \cdot 
        \mathbf{r}} 
\end{eqnarray}
Furthermore, if we consider the case of co-propagating, or nearly
co-propagating, beams along the $z$ axis, these equations read:
\begin{eqnarray} 
    \frac{\partial}{\partial z}\mathcal{E}_{p} & = &\frac{ik_p}{2}\chi_{pp}(\omega_{p})\mathcal{E}_{p}+
    \frac{ik_p}{2}\chi_{pc}(\omega_{p})e^{i\Delta k_z z}\mathcal{E}_{c}^{*}
    \label{probeprop}\\
    \frac{\partial}{\partial z}\mathcal{E}_{c} & = &\frac{ik_c}{2}\chi_{cc}(\omega_{c})\mathcal{E}_{c}+
    \frac{ik_c}{2}\chi_{cp}(\omega_{c})e^{i\Delta k_z z}\mathcal{E}_{p}^{*},
    \label{conjprop} 
\end{eqnarray}
where $\Delta k_z$ is the projection of the geometric phase-mismatch $\Delta
\mathbf{k} = 2 \mathbf{k}_0 - \mathbf{k}_p - \mathbf{k}_c$ on the $z$ axis,
and the conservation of energy imposes the condition $\omega_{p} + \omega_{c}
=2 \omega_{0}$.

In the low pump depletion limit, which is usually experimentally the case, the
pump Rabi frequency $\Omega$ is constant along the vapor cell and these
equations are simply first order coupled linear differential equations.  When
the dynamics is dominated by the cross-terms and no conjugate field is
injected, the probe and conjugate fields grow asymptotically exponentially and
the system behaves like a phase-insensitive amplifier for a probe input
field~\cite{boyd_nonlinear_2008,pooser_low-noise_2009}.
At the quantum level, these cross-terms are responsible for
the creation of quantum correlations between the output probe and conjugate
fields, leading to the production of a two-mode squeezed
state~\cite{reid_demonstration_1989,boyer_entangled_2008}. The larger the gain
of the amplifier, the greater the amount of squeezing.

The full dynamics is more complicated than pure 4WM because of the presence
of the direct terms $\kpp$ and $\kcc$. However the general form of
Eqs.~(\ref{probeprop}) and (\ref{conjprop}) offers us a straightforward
interpretation of $\kpp$ and $\kcc$ in terms of effective linear
susceptibilities for the probe and the conjugate fields. Note that unlike the
usual linear susceptibilities, these effective susceptibilities depend,
nonlinearly, on the pump field. Therefore they give rise to a pump-dependent
complex index of refraction for the probe and the conjugate. The real part
influences the phase-matching of the process, as discussed below.  The
imaginary part translates into absorption.

In the original proposal by Lukin et
al~\cite{lukin_resonant_1998,*lukin_resonant_2000}, the pump beams are
resonant with an atomic transition ($\Delta_1 = \Delta_2 = 0$). This results
in a remarkable situation where the pumps and the twin beams fulfill the
two-photon Raman resonance and enter an electromagnetically-induced
transparency (EIT) condition~\cite{boller_observation_1991}. As a result, they
see a perfectly transparent medium, as witnessed by a vanishing imaginary part
of $\kpp$ and $\kcc$. At the same time, the cross-susceptibilities are
enhanced by the coherence of the hyperfine electronic ground states.  In
theory, it should lead to very efficient 4WM and virtually no absorption, even
for a weak pump.  In practice, the EIT effect in hot vapors is limited because
the Doppler effect, the presence of multiple excited hyperfine levels, and the
finite transit time of the atoms in the laser beams all act to increase the
decoherence rate between the ground states. This causes residual absorption of
the probe field and as a result, only low levels of squeezing have been
observed in this configuration~\cite{van_der_wal_atomic_2003}.

\begin{figure}[htbp]
    \begin{center}
	\includegraphics[width=\linewidth]{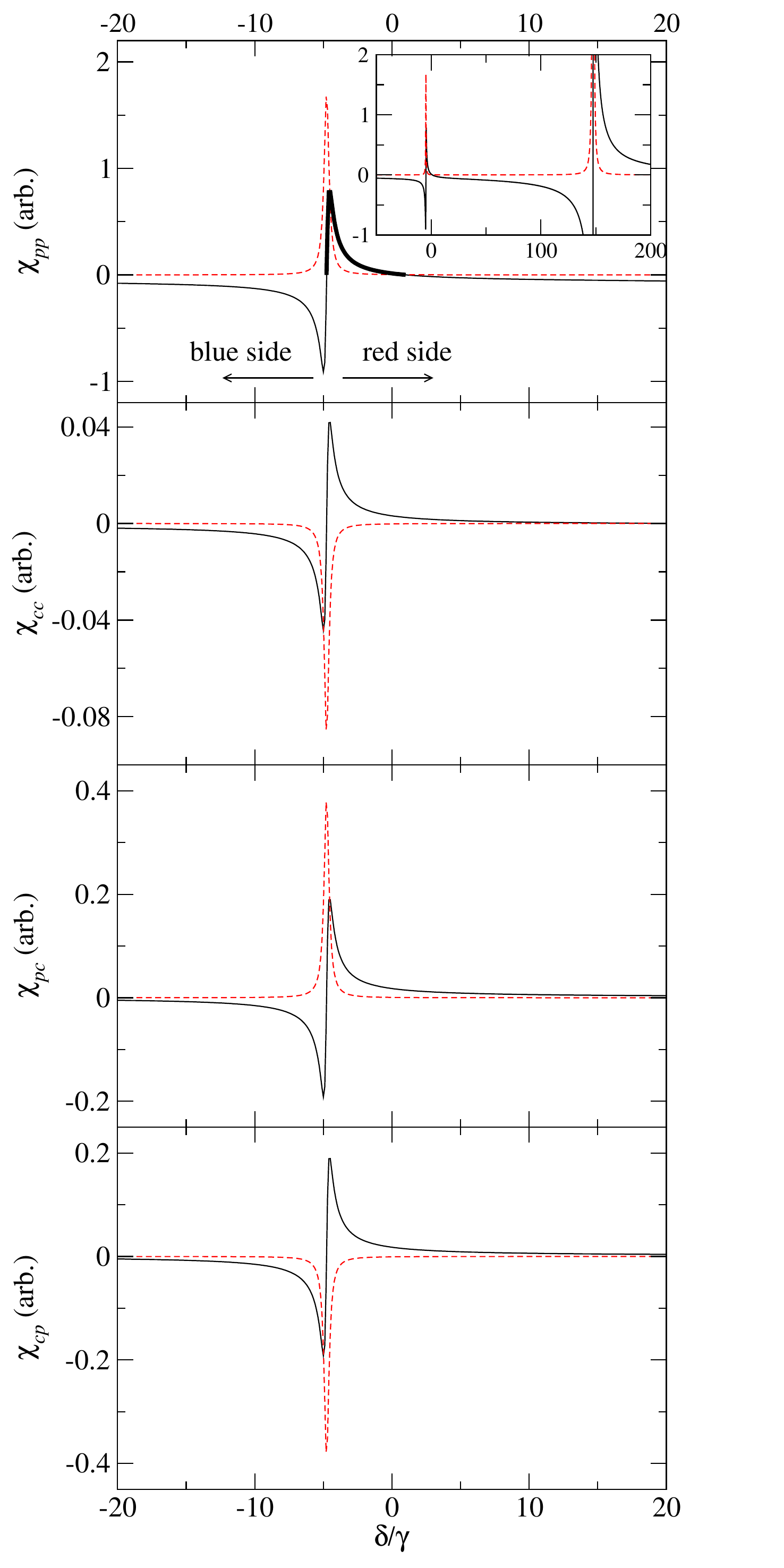}
    \end{center}
    \caption{(color online) The direct and cross susceptibilities for the
    probe and conjugate fields as a function of the two-photon detuning
    $\delta$, varied by changing the probe frequency, expressed in units of
    the excited state decay rate  $\gamma = 2\pi \times 6$ MHz. The solid
    black lines are the real parts, the dashed red lines are the imaginary parts. The
    units on the $y$ axes are arbitrary but identical for all four
    susceptibilities. The resonant Rabi frequency of the pump is $\Omega =
    60\gamma$, the detuning of the pump is $\Delta_1 = 140 \gamma$, and the
    decoherence rate of the excited state is estimated (see below) to be
    $\gamma_c = 0.2\gamma$. 
    %The 4WM process is the feature close to
    %$\delta=0$. 
    The feature at $\delta = 150 \gamma$, visible on $\kpp$ in the
    insert, corresponds to the
    one-photon resonance of the probe from the $F=2$ hyperfine ground state.
    The region of positive $\Re(\chi_{pp})$ around the 4WM resonance is
    indicated with a thicker line.}
    \label{fig:chi}
\end{figure}

In contrast, most recent squeezing experiments in hot atomic vapors operate at
large detuning $\Delta_1/2\pi$, typically 0.5 to 1 GHz, and at larger pump
power~\cite{mccormick_strong_2007,glorieux_quantum_2011,jasperse_relative_2011,
qin_compact_2012}.  In these conditions, the susceptibilities take a different
form from the resonant case.  This off-resonant form is depicted in
Fig.~\ref{fig:chi}, for typical experimental parameters.

The first important property of these susceptibilities is that the 4WM resonance
is shifted from the bare two-photon resonance ($\delta = 0$) by the light shift
created by the more resonant pump, on the $5\mathrm{S}_{1/2}(F = 2)
\rightarrow 5\mathrm{P}_{1/2}$ transition. For typical
experimental parameters, the shift is of the order of $-5 \gamma$.  In the
rest of the paper, we call ``blue side'' of the 4WM resonance the range of
detunings $\delta$ for which the frequency of the probe is above the 4WM
resonance. The other side of the 4WM resonance is the ``red side'' (see
Fig.~\ref{fig:chi}).  

\begin{figure}[htb]
    \begin{center}
	\includegraphics[height=.95\linewidth,angle=-90]{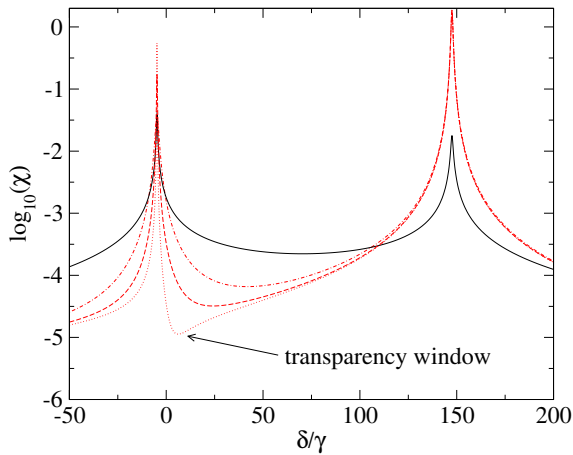}
    \end{center}
    \caption{4WM coupling  $|\chi_{pc}|$ (solid black line) and probe
    absorption $\Im(\chi_{pp})$ (dashed red line) as a function of the two
    photon detuning $\delta$, for  $\gamma_c = 0.2\gamma$.
    The probe absorption is also shown for $\gamma_c = \gamma$ (dashed-dotted
    line) and for $\gamma_c = 0.02\gamma$ (dotted line).  In the latter case,
    the transparency window around $\delta = 0$ is visible.}
    \label{fig:logchi}
\end{figure}

The second property is that the imaginary part of $\kpp$ (dashed line) is
maximum at the 4WM resonance, due to Raman absorption. It is therefore not a
good place to observe 4WM because the medium is essentially opaque for the
probe at this detuning. On the other hand, $\Im (\kpp)$ decays much faster
than the magnitude of $\kpc$ when moving away from resonance, therefore there
is a range of $\delta$ on each side of the 4WM resonance where $\kpc$
still exhibits a substantial magnitude while $\Im (\kpp)$ has
almost completely vanished (see Fig.~\ref{fig:logchi}). These regions of the
two-photon detuning are better places to observe quantum effects.

The third important property is the behavior of the real part of $\kpp$ (solid
line), which is effectively responsible for the index of refraction for the
probe. Around the 4WM resonance, $\Re (\kpp)$ is the sum of a dispersive
feature resulting from the 4WM coupling and the off-resonance negative
susceptibility resulting from the one-photon transition between ground state
and excited state. These competing terms lead to a cancellation of $\Re
(\kpp)$ around the bare two-photon resonance ($\delta = 0$).

It is legitimate to wonder if it would not be possible to take advantage of
the coherence between the ground states to reduce the probe absorption through
the EIT phenomenon, as envisioned in the original proposal. Theoretically, for
a very low value of the ground state decoherence rate $\gamma_c$, it is indeed
possible to observe the transparency window at the bare two-photon resonance,
as shown in Fig.~\ref{fig:logchi}.  In practice, such low decoherence rates
are not achieved due to imperfections associated with hot vapors. Experimental
results reported below and in
Refs~\onlinecite{boyer_ultraslow_2007,glorieux_double-_2010} are compatible with a
higher value $\gamma_c \simeq 0.2 \gamma$, for which there is no marked
transparency window. In spite of the lack of efficient EIT, the off-resonant
value of $\Im (\kpp)$ is small enough compared to the cross-coupling to ensure
efficient 4WM, and experimental set-ups do not require special precautions
with regards to decoherence sources such as stray magnetic fields.

Note that these considerations do not apply to the conjugate field, which is
much further detuned from resonance than the probe. The direct susceptibility of
the conjugate $\kcc$ is substantially smaller than the other susceptibilities
and can be regarded as zero in practice.

Now that adequate ranges of $\delta$ have been identified for which probe
losses are negligible compared to 4WM amplification, the question is whether
there is a geometric configuration of the light fields for which 4WM is phase
matched. To answer this question, we recall from
Ref.~\onlinecite{lukin_resonant_1998,*lukin_resonant_2000} the solutions to the
propagation equations (\ref{probeprop}) and (\ref{conjprop}), for a seed
$\mathcal{E}_s$ on the input probe and no input conjugate:
\begin{eqnarray} \mathcal{E}_p & = & \mathcal{E}_s \exp (\delta aL)
    \left[\cosh(\xi L) + \frac{a}{\xi}\sinh(\xi L) \right] \label{prop_p} \\
    \mathcal{E}_c^* & = & \mathcal{E}_s \exp (\delta aL) \frac{a_{cp}}{\xi}
    \sinh(\xi L) \label{prop_c} \end{eqnarray}
where $L$ is the length of the medium, $a_{pj}=ik_p\chi_{pj}/2$,
$a_{cj}=ik_c\chi_{cj}^*/2$, $\delta a = (a_{pp} - a_{cc} + i\Delta k_z)/2$, $a
= (a_{pp} + a_{cc} - i\Delta k_z)/2$, $\xi = \sqrt{-a_{pc}a_{cp} + a^2}$, and
$j = p,c$.  We define the probe and conjugate intensity gains $g_p$ and $g_c$
as $|\mathcal{E}_p|^2 = g_p |\mathcal{E}_s|^2$ and $|\mathcal{E}_c|^2 = g_c
|\mathcal{E}_s|^2$.

\begin{figure}[htb]
    \begin{center}
	\includegraphics[width=.9\linewidth]{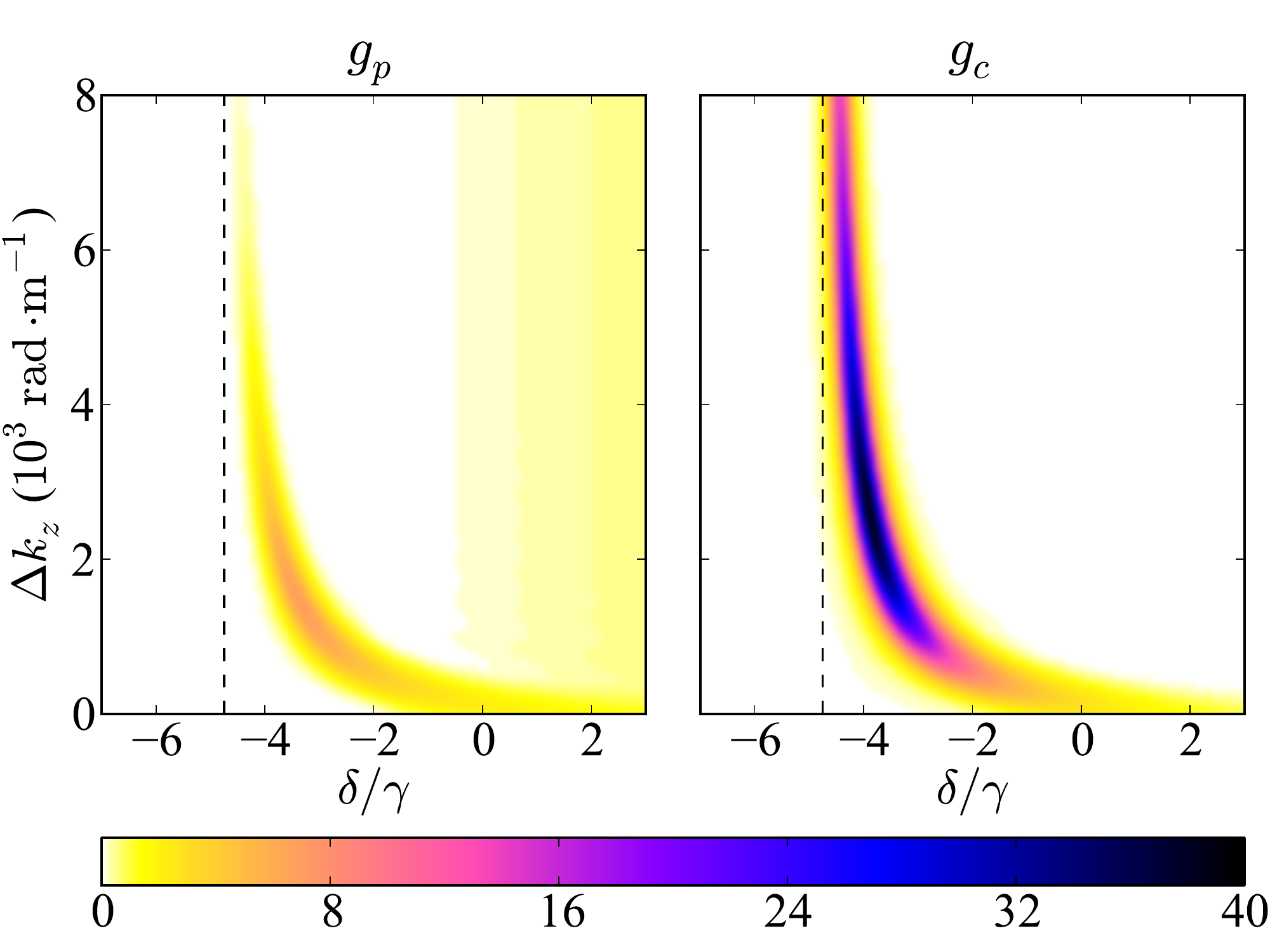}
    \end{center}
    \caption{Theoretical output probe and conjugate gains $g_p$ and $g_c$ as a
    function of the two-photon detuning and the geometrical phase mismatch.
    Here the decoherence rate is $\gamma_c = 0.5\gamma$, the atom density is
    $\mathcal{N} = 3\times 10^{12}\, \mathrm{cm^{-3}}$, the length of the
    medium is $L = 12.5\, \mathrm{mm}$ and the pump Rabi frequency is $\Omega
    = 60\gamma$.  The dashed lines indicate the position of the 4WM
    resonance.}
    \label{fig:match}
\end{figure}

From these expressions, we plot in Fig.~\ref{fig:match} the
probe and conjugate gains as a function of $\delta$ and the geometric phase
mismatch $\Delta k_z$.  One can see that for $\Delta k_z \simeq 0$, the
maximum gain is obtained around $\delta \simeq 0$, for both the probe and the
conjugate. When $\Delta k_z$ increases, the gain on both the probe and the
conjugate increases while the gain resonance moves towards the 4WM resonance
(shown with the vertical dashed line). At larger $\Delta
k_z$, the gain resonance comes asymptotically within $\gamma$ of the 4WM
resonance, the probe intensity drops while the conjugate intensity keeps
increasing. Finally, at large $\Delta k_z$, the conjugate intensity also
drops.

The position of the gain resonance is the result of the 4WM phase-matching (or
absence thereof), and is influenced by the effective index of refraction seen
by the probe and the conjugate as follows. Since $\Re(\kcc)$ is much smaller
than the other susceptibilities, the conjugate effectively propagates in a
medium of index 1.  The situation is different for the probe, for which the
index of refraction changes sign at the 4WM resonance and at $\delta \simeq
0$, as indicated in Fig.~\ref{fig:chi}.  The change in sign of $\Re(\kpp)$ at
$\delta \simeq 0$ means that the probe experiences an effective index of
refraction, $n_p$, smaller than 1 for $\delta \gtrsim 0$ and larger than 1
between $\delta \lesssim 0$ and the 4WM resonance. We assume that the index of
refraction experienced by the pump, $n_0$, is close to 1 since the pump tends
to optically pump the atoms towards the ground state of the off-resonant
transition $5\mathrm{S}_{1/2}(F = 3) \rightarrow 5\mathrm{P}_{1/2}$. This
assumption will be refined later.

\begin{figure}[htb]
    \begin{center}
	\includegraphics[width=0.8\linewidth]{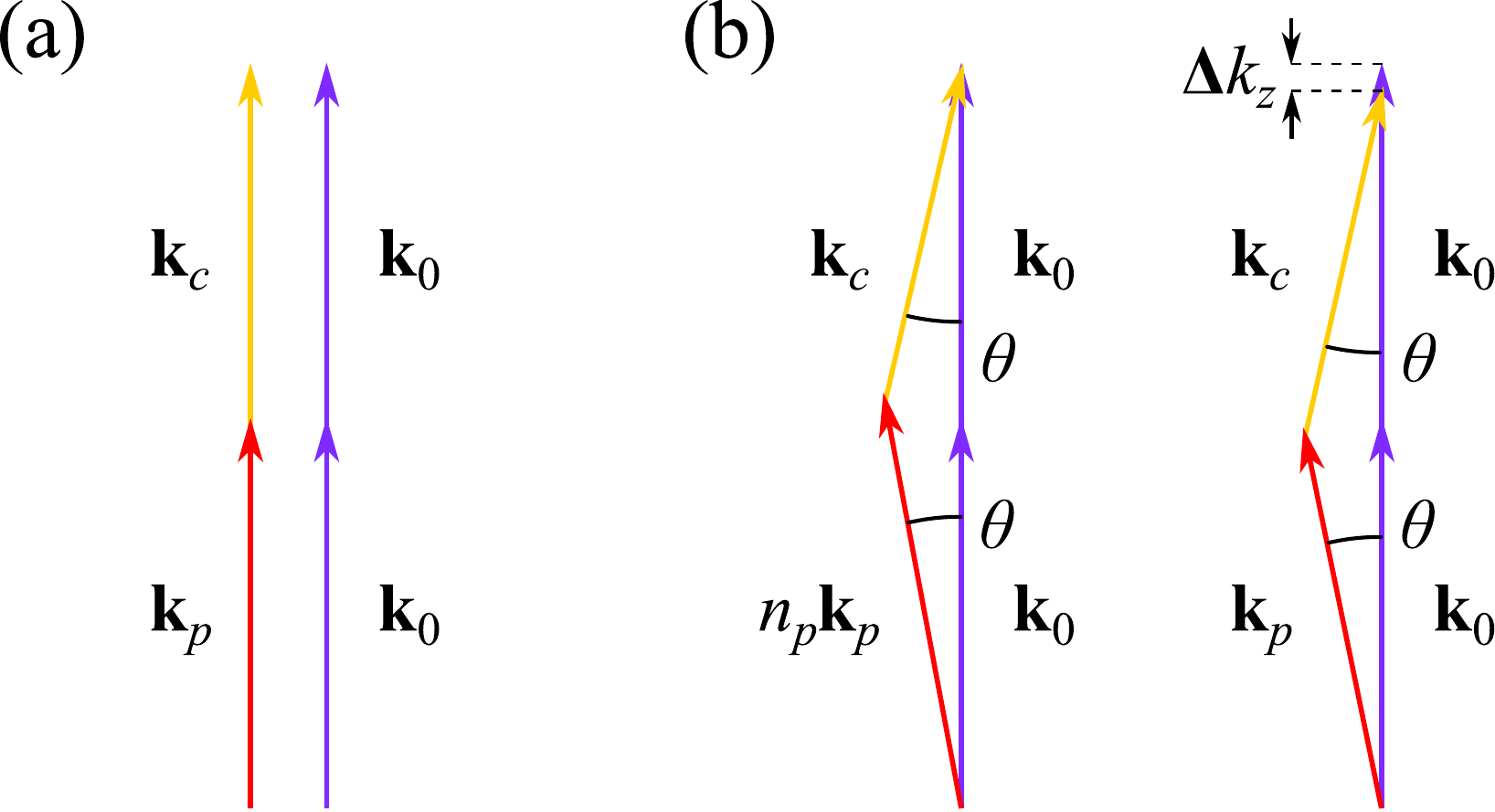}
    \end{center}
    \caption{(a): configuration where the geometric phase-matching condition
    is fulfilled ($\Delta k_z = 0$); (b): configuration where the effective
    phase-matching condition for an effective index of refraction of the probe
    $n_p \gtrsim 1$ is fulfilled. In this case, there is a necessary geometric phase
    mismatch ($\Delta k_z > 0$). The wave-vectors in vacuum $\mathbf{k}_0$,
    $\mathbf{k}_p$ and $\mathbf{k}_c$, for the pump, probe and conjugate
    respectively, all have nearly the same magnitude. Energy conservation
    ensures that $k_p + k_c = 2 k_0$.}
    \label{fig:matching}
\end{figure}

The geometric phase-matching condition $\Delta k_z = 0$ is the phase-matching
condition in free space. It is fulfilled only when the beams are rigorously
co-propagating, as shown in Fig.~\ref{fig:matching}(a).  For the process to be
efficient, the effective phase-matching condition must be fulfilled:
\begin{equation} 2 \mathbf{k}_0 - n_p \mathbf{k}_p - \mathbf{k}_c = 0,
    \label{phase-matching} \end{equation}
as shown in Fig.~\ref{fig:matching}(b). This condition is identical to the
geometric phase-matching condition ($\Delta k_z = 0$) only when $n_p = \sqrt{1
+ \Re (\kpp)} = 1$, which occurs around $\delta = 0$ (Fig.~\ref{fig:chi}).
When $n_p > 1$, the effective phase-matching condition (\ref{phase-matching})
imposes $\Delta k_z > 0$, which corresponds to having a finite angle $\theta$
between the pump and the probe and conjugate~\footnote{As the angle
between the pump and the probe, $\theta$, is small and $n_p$ is close to 1, the angle
between the pump and the conjugate is very close to $\theta$.}.
%For $\Delta k_z > 0$, which corresponds to having a finite angle $\theta$
%between the pump and the probe and conjugate, the effective phase matching
%condition requires $n_p > 1$. 
This occurs on the red side of the 4WM resonance, for $\delta \lesssim 0$ (see
Fig.~\ref{fig:chi}). As $\theta$ increases, the gain resonance is shifted
towards higher values of $\Re(\kpp)$ and therefore towards the 4WM resonance.
A negative geometric phase mismatch $\Delta k_z$ cannot be fulfilled.  For
this reason, no effective phase matching can happen on the blue side of the
4WM resonance, where $n_p < 1$.

Close to the 4WM resonance, the increase in $\Im(\kpp)$ accounts for the
reduction in probe power with respect to the conjugate power, seen in
Fig.~\ref{fig:match}. In this region, a high level of probe absorption coupled
to a large 4WM gain still produces a strong conjugate output. At larger angle
$\theta$, the effective phase-matching condition requires a value of $\delta$
so close to the 4WM resonance that the high probe loss prevents the 4WM from
happening at all.

\section{Experimental verification} 
\label{experiment}

\begin{figure}[htb]
    \begin{center}
	\includegraphics[width=.9\linewidth]{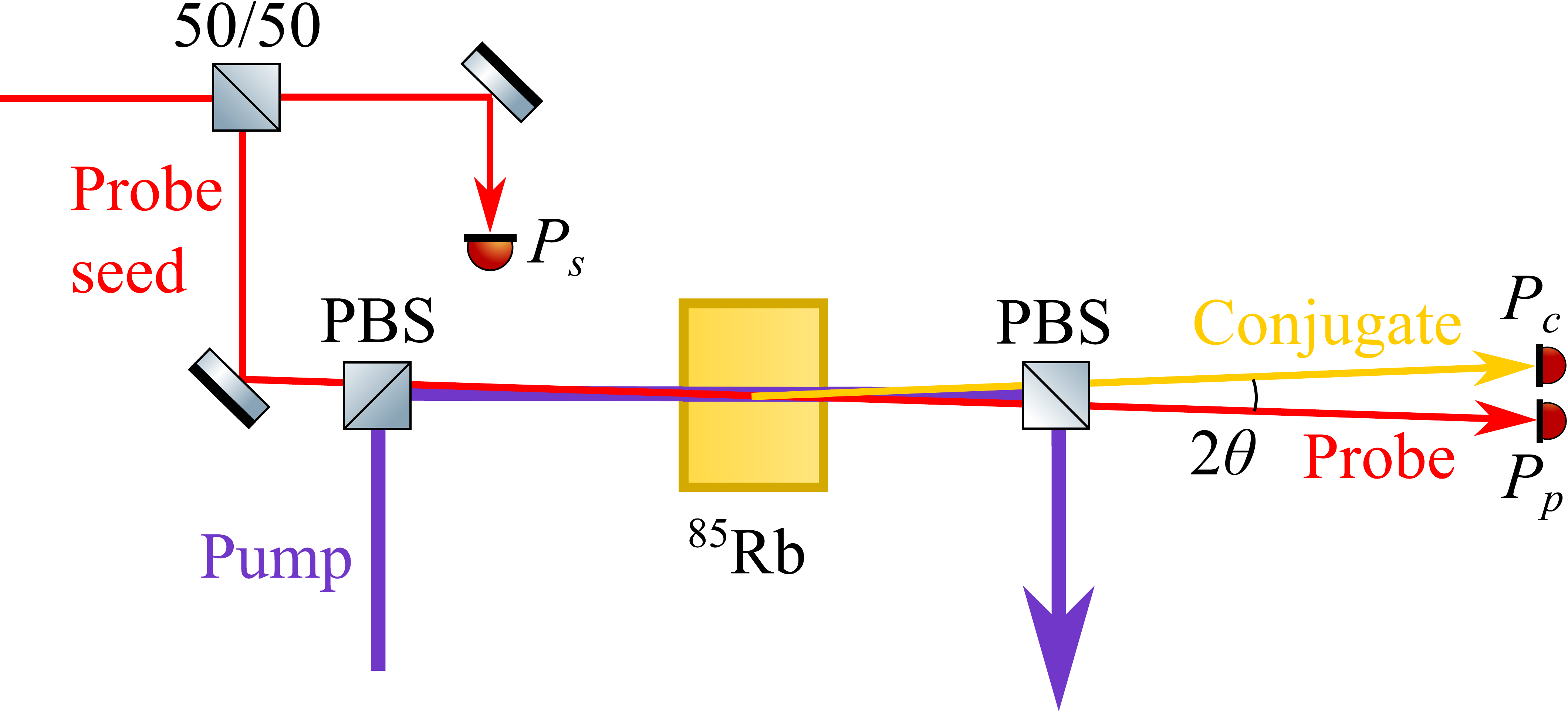}
    \end{center}
    \caption{The Rb cell is pumped with a bright pump beam, in blue, and seeded
    at an angle $\theta$
    with a probe of power $P_s$, in red. Emitted is an amplified probe of
    power $P_p$, also red, and a conjugate of power $P_c$, in yellow. The
    three powers, $P_s$, $P_p$ and $P_c$, are monitored by photodiodes while
    the probe frequency is scanned. PBS: polarizing beamsplitter.}
    \label{fig:Apparatus}
\end{figure}

In order to verify the theoretical predictions, a test was performed, as shown
in Fig.~\ref{fig:Apparatus} in which a $750$~mW pump laser of beam waist
$0.9$~mm drives the D1 line at 795~nm in a 12-mm-long cell of $^{85}$Rb
vapor, heated and temperature stabilized at $\sim 110^\circ$C. A seed beam at
the probe frequency is produced by diverting a fraction of the pump through an
AOM operating at $\omega_\mathrm{HF}/2 \simeq 2\pi \times 1.5$~GHz in a double-pass
arrangement. This seed beam, of power $10-20$~$\mu$W and waist $0.4$~mm, then
intersects with the pump inside the cell at a small angle $\theta$. The probe
and conjugate beams are perpendicularly polarized with respect to the pump
which is rejected at the output with a polarizing beam splitter.
From the measured input seed power $P_s$, output probe power $P_p$, and output
conjugate power $P_c$, the probe and conjugate gains ($g_{p}
= P_p/P_s$ and  $g_{c} = P_c/P_s$, respectively) are obtained. 

\begin{figure}[htb]
    \begin{center}
	\includegraphics[width=.9\linewidth]{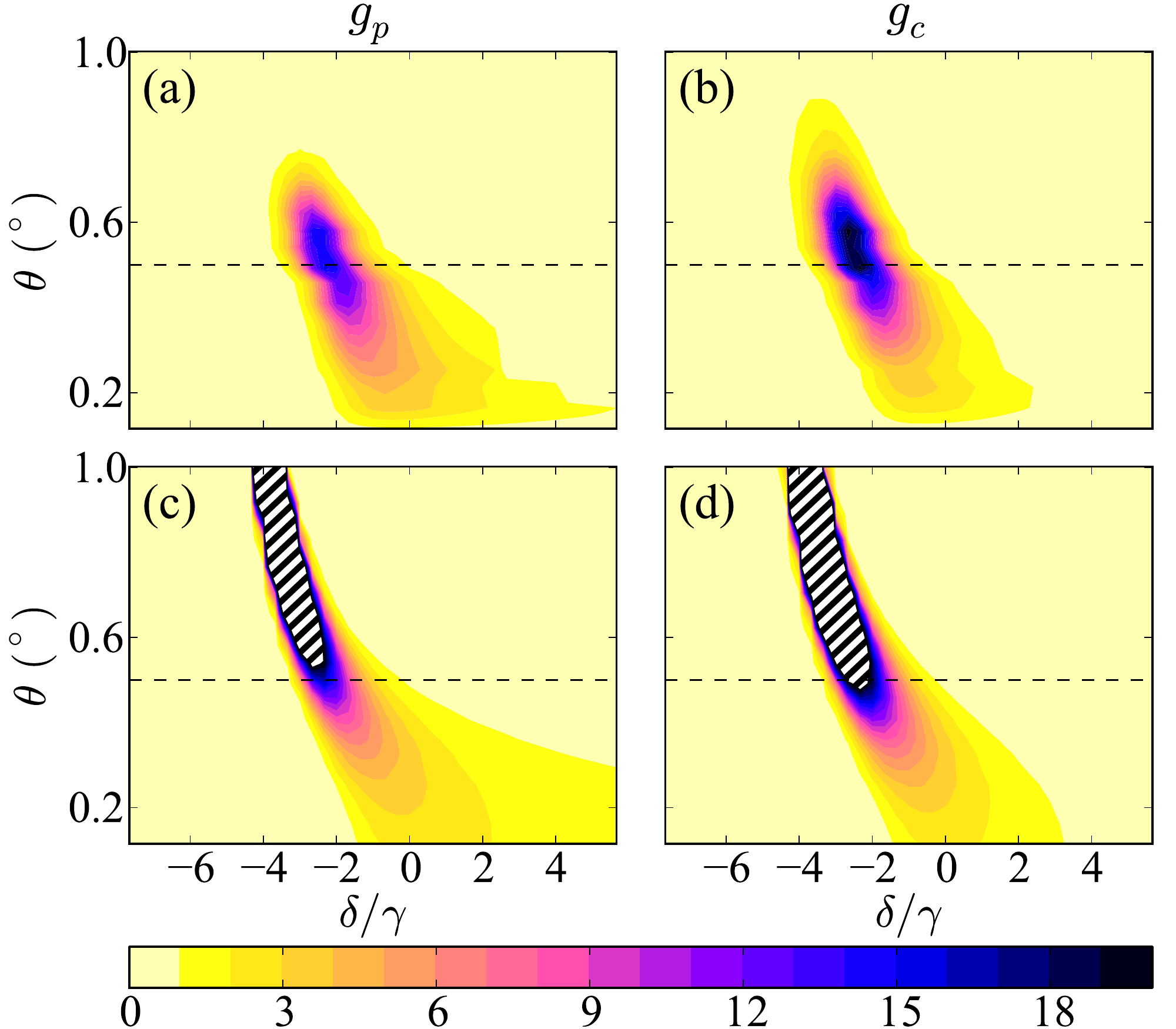}
    \end{center}
    \caption{Probe gain $g_p$ and conjugate gain $g_c$ as a function of
    the two-photon detuning and the probe-pump angle. Top: experimental data. 
    Bottom: fit to the experimental data for the angles between 0.1$^\circ$
    and 0.5$^\circ$, that is to say for the region below the dashed line. The
    hatched regions represent values outside the color scale. The
    theoretical peak gains for the probe and the conjugate are 350 and 1100 
    respectively.}
    \label{fig:fit}
\end{figure}

A dichroic-atomic-vapor laser lock~\cite{corwin_frequency-stabilized_1998} is
in place to regulate the pump frequency thus maintaining a constant
$\Delta_1$.  The angle of intersection $\theta$ between the pump and probe is
set by manually adjusting a pair of input mirrors, and $\delta$ is adjusted by
changing the AOM drive frequency.  For a selection of values of $\theta$,
$\delta/2\pi$ is swept over a range of typically $60$~MHz, where noticeable gains
$g_p$ and $g_c$  are observed. The angular range used, from 0$^\circ$ to
1$^{\circ}$, satisfies the condition that there must be a full overlap of the
beams over the cell length (this is achievable for $\theta$ up to
$5^{\circ}$).  Figures \ref{fig:fit}(a, b) show the contour plots of $g_p$ and
$g_c$ as a function of $\delta$ and $\theta$.

The general features of figure \ref{fig:match} are reproduced, including the
shift of the gain peaks towards the 4WM resonance when $\theta$ is increased,
as well as the crossover between probe power and conjugate power. The
crossover is the value of $\theta$ for which the peak conjugate power is equal
to the peak probe power. The main discrepancy between the experimental data
and the theoretical prediction is the measured drop in probe and conjugate
power for $\theta > 0.6^\circ$. This leads to peak gains for both the probe
and the conjugate which are well below the theoretical prediction. There are
two main reasons for this behavior.  Firstly, for a finite $\theta$, the
Doppler effect due to the thermal motion of the atoms does not cancel between
the pump and the twin beams. For $\theta = 1^\circ$, the residual Doppler
effect on $\delta$ reaches $2\gamma$, which is roughly the width of the gain
peak itself. This results in a broadening of the gain resonance at larger
angles $\theta$, when compared to the resonance given by the theoretical
model, which is visible in Fig.~\ref{fig:fit}.

Secondly, and perhaps more importantly, when $\theta$ is increased past
0.5$^\circ$, the probe beam is subject to a strong effective cross-Kerr
interaction with the pump around the gain resonance. This is due to the fact
that as the gain peak moves closer to the 4WM resonance the effective index of
the probe is resonantly enhanced, as shown by the behavior of $\Re (\kpp)$ in
Fig.~\ref{fig:chi}. As a result, the transverse intensity variation of the
pump realizes a strong lens for the probe, causing it to emerge from the cell
with an angle of divergence comparable to or larger than $\theta$
itself~\footnote{The full Kerr action of the pump on the probe is the compound
effect of the change of magnitude of the index of refraction at resonance and
the frequency shift of the resonance due to the light shift caused by the
pump.}.

The good agreement between the theoretical model and the measurements at
angles where the probe focusing is negligible gives us the opportunity to
extract the values of those parameters which are not otherwise easily
obtained, in particular the pump Rabi frequency $\Omega$, the atom density
$\mathcal{N}$, and most importantly the decoherence rate $\gamma_c$.
Moreover, as explained in the appendix, we can also introduce the
effect of the index of refraction of the pump on the phase-matching condition
by replacing the geometric phase mismatch with $\Delta k_z = 2n_0 k_0 - k_p
\cos \theta - k_c \cos \theta$, where $n_0 = 1 - \varepsilon$ is the pump
index of refraction.  By fitting the model on the data, as shown in
Figs.~\ref{fig:fit}(c, d), we find that $\Omega = 60\gamma$, $\mathcal{N} =
2.8 \times 10^{12}\, \mathrm{cm^{-3}}$, $\gamma_c = 0.2 \gamma$, and
$\varepsilon = 6.5 \times 10^{-6}$. For this data set, the pump detuning was
determined to be $\Delta_1 = 140\gamma$ by calibrating the position of the 4WM
gain against a Rb spectroscopy spectrum.  As expected, the index of refraction
for the pump on the blue side of the atomic resonance is smaller than one.

One can check that the parameters extracted from the fit are broadly
consistent with the estimated experimental conditions. Our pump beam
parameters lead to a peak intensity of $60\, \mathrm{W\cdot cm^{-2}}$, which
results in a resonant Rabi frequency $\Omega = 80 \gamma$ for a mean electric
dipole $d = 1.47 \times 10^{-29}\, \mathrm{C\cdot
m}$~\cite{steck_rubidium_2012}.  From the vapor pressure data summarized in
Ref.~\onlinecite{steck_rubidium_2012}, the number density of $^{85}$Rb at
$100^\circ$C is $\mathcal{N} \simeq 4 \times 10^{12}\, \mathrm{cm^{-3}}$.
Finally the index of refraction for the pump, evaluated for ground state
populations of 6\% in the lower hyperfine state and 94\% in the upper
hyperfine state, as given by Eqs.~(\ref{pop1}--\ref{pop4}), and for the
electric dipole value given above, is $n_0 = 1 - 1.6 \times 10^{-5}$.

\section{Doppler broadening}
\label{doppler}

The model developed above does not take into account the Doppler broadening
caused by the thermal motion of the atoms in the cell. The good agreement
between the model and the experimental data suggests that considering only
average values of the single-photon detunings $\Delta_1$ and $\Delta_2$
captures most of the physics at play. However, considering that the probe is
typically tuned to the edge of the Doppler profile, it is legitimate to wonder
what level of absorption this causes. It turns out that although the EIT has a
limited impact, the pump field is highly saturating at resonance and causes a
wide Autler-Townes splitting for those atoms resonant with the probe field.
This renders even the resonant part of the atomic vapor highly transparent for
the probe as long as the optical depth is not too large. In practice,
noticeable levels of squeezing can be observed for $\Delta_1/2\pi$ as low as
500~MHz~\cite{mccormick_strong_2008}, which is well inside the Doppler profile
at our operational temperature.

\begin{figure}[htb]
    \begin{center}
	\includegraphics[width=.9\linewidth]{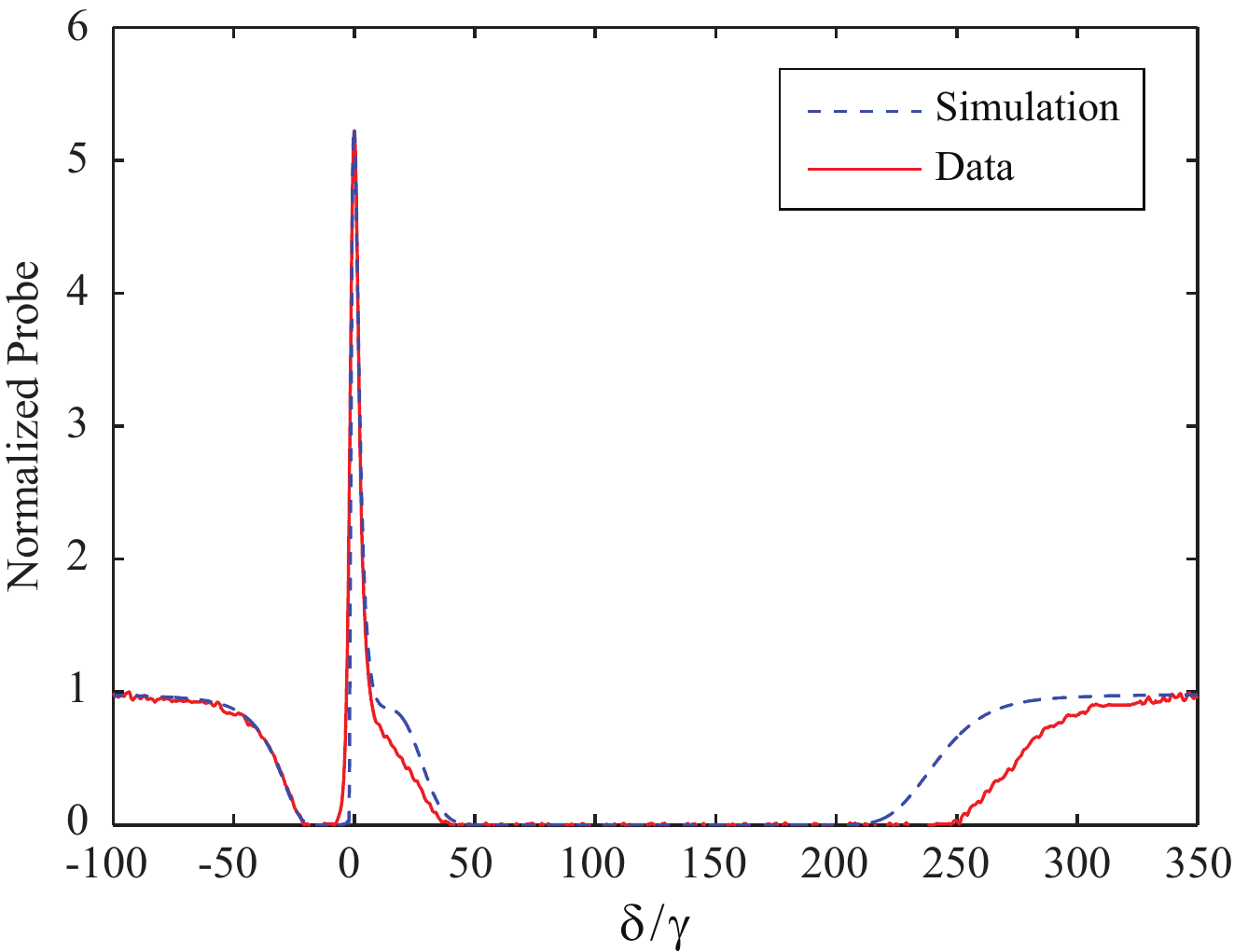}
    \end{center}
    \caption{Probe gain as a function of the two-photon detuning at low angle
    $\theta$ and fit to the
    data including the Doppler broadening in the model. The discrepancy in the
    width of the Doppler-broadened single-photon absorption dip is caused by
    the presence of 2 excited hyperfine levels separated by $60\gamma$. The model
    only considers two excited states which are degenerate in energy
    (Fig.~\ref{fig:lambda2}).}
    \label{fig:doppler}
\end{figure}

In order to verify these assumptions, we extended the above model to include
the full Doppler distribution of detunings. The result, fitted to the gain
curve of the probe as a function of $\delta$, is shown in Fig.~\ref{fig:doppler}.
The small discrepancy in the width of the single-photon resonance is due to
the fact that the model does not include the hyperfine structure of the
excited state, whose main effect is to broaden the apparent Doppler profile.

\section{Optimizing for quantum noise reduction}
\label{quantum}

The observation of large levels of intensity-difference quantum noise
reduction requires a near-perfect phase-insensitive amplifier with a gain
of at least a few units. In the case of 4WM in a hot atomic vapor, this
means that the absorption of the twin beams must be kept to a minimum while
ensuring that efficient 4WM can take place. As pointed out previously, the
two-photon detuning which fulfills those two conditions is $\delta \simeq 0$,
and not the 4WM resonance in itself. This is firstly because away from the 4WM
resonance the probe susceptibility responsible for the absorption drops faster
than the cross-susceptibility responsible for the 4WM gain, and secondly
because the ground state coherence reduces the probe absorption at
that detuning~\footnote{It is not exactly $\delta = 0$ due to the light shift
created by the pump beam on the $5\mathrm{S}_{1/2}(F = 3) \rightarrow
5\mathrm{P}_{1/2}$ transition.}.

From the previous discussion, at a detuning $\delta \simeq 0$, effective phase
matching of the 4WM process requires geometric phase-matching $\Delta k_z =
0$.  When corrected for the index of refraction for the pump, this condition
corresponds to the introduction of a small angle $\theta \ll 1^\circ$ between
the pump and the probe beams.  This is indeed how the best levels of squeezing
have been experimentally observed~\cite{mccormick_strong_2008,glorieux_quantum_2011,
jasperse_relative_2011,qin_compact_2012}. The production of squeezed light by
non-degenerate 4WM in a hot vapor therefore benefits from two favorable
circumstances. First, the optimum angle $\theta$ is small enough that the
residual Doppler effect acting on the nearly co-propagating beams is much
smaller than the width of the gain peak.  Second, the same angle is large
enough that the beams participating in the 4WM can be spatially separated at
the output of the vapor cell.

It is worth noting that in certain conditions, a small amount of loss on the
probe beam can be beneficial. For instance, for a probe seed containing a
large amount of classical noise, it is useful to have twin beams of equal
powers in order to ensure proper rejection of the classical noise in the
balanced detection~\cite{mccormick_strong_2008}\footnote{We consider here only
the case where the detector is perfectly balanced, with equal electronic gains
on both inputs.}. In particular, the existence of points in the $(\delta,
\theta)$ parameter space where the probe and conjugate powers are perfectly
balanced has allowed the detection of 8~dB of intensity-difference squeezing
at frequencies as low as 2.5~kHz, despite the presence of substantial
technical noise on the input probe~\cite{mccormick_strong_2008}.

\section*{Conclusion}

We have shown that it is possible to phase-match 4WM in a hot atomic vapor so that
absorption is reduced to a level where quantum effects could manifest themselves.
Most reports of large squeezing generated by non-degenerate forward 4WM
in hot vapors to date have indeed used a similar arrangement as the one presented
here, with nearly identical atomic and beam parameters. Furthermore, the model
we have developed appears very accurate for small angles between beams, even
in its simple form neglecting both the Doppler effect and the Zeeman
sub-structure. It allows the robust extraction of parameters which would be
difficult to determine by direct measurement, such as the decoherence rate of
the ground states.

By providing further
insight into the mechanism of the process, the present results may lead to the
realization of different configurations or different regimes amenable to the
production of interesting quantum states of light, such as multi-spatial-mode
phase-sensitive
amplifiers~\cite{corzo_multi-spatial-mode_2011,*corzo_noiseless_2012}.

We acknowledge useful discussions with Colin McCormick, Ennio Arimondo and
Paul Lett, and we thank Etienne Pertreux for his help in the early stages of
this project. This research was supported by the Engineering and Physical
Sciences Research Council grants EP/E036473/1 and
EP/I001743/1.

\appendix*

\section{Expressions for the susceptibilities}
\label{appendix}

\subsection{Derivation of the susceptibilities for the probe and the
conjugate}

In this appendix we derive the dynamics of the double-lambda configuration
described in the paper.  The response of an atomic system to an optical field
is determined by the polarization of the medium, which acts as the driving
term in the wave equation.  For a medium that consists of non-interacting
particles, such as a dilute atomic vapor, the polarization of the medium is of
the form \cite{boyd_nonlinear_2008}
\begin{equation*}
    \textbf{P}=\mathcal{N}\langle\hat{\textbf{d}}\rangle,
\end{equation*}
where $\mathcal{N}$ is the number density of the atomic medium and
$\hat{\textbf{d}}$ is the atomic dipole moment operator.  The polarization of
the medium can be written in terms of the atomic eigenstates, such that it
takes the form
\begin{equation*}
    \textbf{P}=\mathcal{N}\sum_{n,m}\textbf{d}_{mn}\sigma_{nm}e^{-i\omega_{f\!,nm}t},
\end{equation*}
where the sum is over all the involved atomic transitions, $\omega_{f\!,nm}$ is
the frequency of the field that couples the transition between levels $n$ and
$m$, $\sigma_{nm}$ is the density matrix element between levels $n$ and
$m$ in a rotating frame at frequency $\omega_{f\!,nm}$, and $\textbf{d}_{mn}$ is the dipole matrix element between levels $n$ and
$m$.  For an isotropic medium, the polarization of the
atomic medium at a particular frequency is  given by
\begin{equation*}
    \textbf{P}(\omega_{f\!,nm}) = \mathcal{N}\textbf{d}_{mn}\sigma_{nm}.
\end{equation*}
Thus, the response is completely determined by the atomic coherence of the
corresponding transition.

The equations of motion for the density matrix elements in the rotating frame
can be shown to be of the form~\cite{boyd_nonlinear_2008}
\begin{widetext} 
\begin{eqnarray*} 
    \dot{\sigma}_{nm}&=&
    i(\Delta_{nm}-\gamma_{nm})\sigma_{nm}+\frac{i}{\hbar}\sum_{\nu}\left[\textbf{d}_{n\nu}\cdot
    \textbf{E}(\mathbf{r}, t)\sigma_{\nu m}e^{-i(\omega_{f\!,\nu
    m}-\omega_{f\!,nm})t}%\right.\nonumber\\
    %&& \left.
    -\sigma_{n \nu}\textbf{d}_{\nu m}\cdot \textbf{E}(\mathbf{r},
    t)e^{-i(\omega_{f\!,n\nu}-\omega_{f\!,nm})t}\right]
    \quad\mbox{for $n\neq m$} \nonumber\\
    \dot{\sigma}_{nn}&=& \frac{i}{\hbar}\sum_{\nu}\left[\textbf{d}_{n\nu}\cdot
    \textbf{E}(\mathbf{r}, t)\sigma_{\nu n}e^{-i\omega_{f\!,\nu n}t}-\sigma_{n
    \nu}\textbf{d}_{\nu n}\cdot \textbf{E}(\mathbf{r}, t)e^{-i\omega_{f\!,n
    \nu}t}\right]%\nonumber\\
    %&&
    +\sum_{E_{m}>E_{n}}\Gamma_{nm}\sigma_{mm}-\sum_{E_{m}<E_{n}}\Gamma_{mn}\sigma_{nn},
    \label{densmatRWA} \end{eqnarray*} \end{widetext}
where $\Delta_{nm}$ is the detuning of the field at frequency $\omega_{f\!,nm}$
from the transition between levels $n$ and $m$ , $\Gamma_{mn}$ is the
population decay rate from level $n$ to level $m$,
$\gamma_{nm}=(\Gamma_{n}+\Gamma_{m})/2+\gamma_{nm}^{c}$ is the dipole
dephasing rate, $\Gamma_{n}$ is the total decay rate out of level $n$, and
$\gamma_{nm}^{c}$ is the dipole dephasing rate due to any other source of
decoherence.

\begin{figure}[tbh]
    \begin{center}
	\includegraphics[width=.5\linewidth]{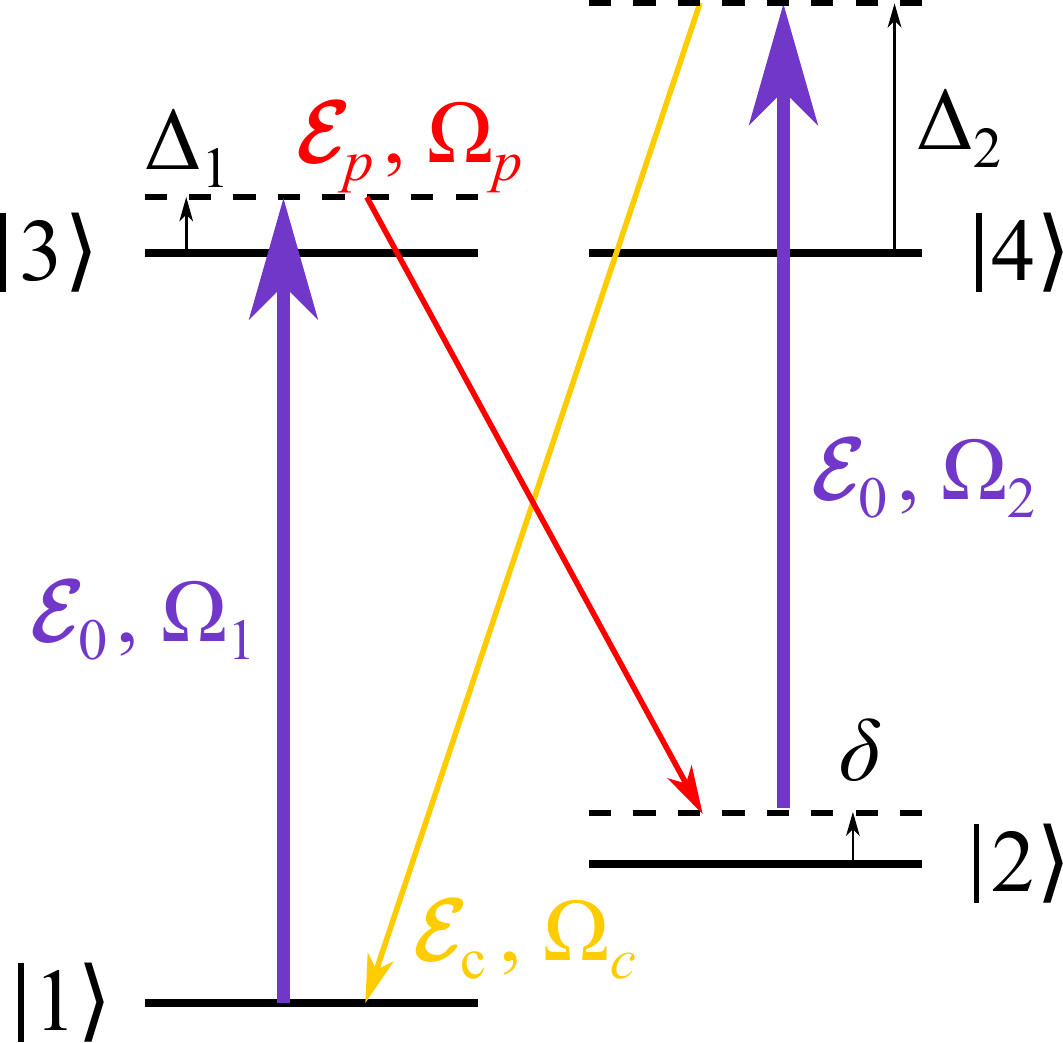}
    \end{center}
    \caption{Double-lambda scheme with a single pump field $\mathcal{E}_{0}$.
    States $|3\rangle$ and $|4\rangle$ are orthogonal linear combinations of
    magnetic states of the excited hyperfine levels.
    }     
    \label{fig:lambda2}
\end{figure}

We now specialize to our 4WM process in the double-lambda configuration, shown
in Figs.~\ref{fig:lambda} and \ref{fig:lambda2}, with
a single pump field, $\mathcal{E}_{0}$. For this case the total electric field is of the form
\begin{eqnarray*} 
    \textbf{E}(\mathbf{r}, t)&=&\mathcal{E}_{0} e^{i(\mathbf{k}_0\cdot 
    \mathbf{r} - \omega_{0}t)}{\boldsymbol
    \epsilon}_{0} +\mathcal{E}_{c}e^{i(\mathbf{k}_c\cdot
    \mathbf{r}- \omega_{c}t)}{\boldsymbol \epsilon}_{c} \nonumber\\
    &&+\mathcal{E}_{p}e^{i(\mathbf{k}_p\cdot \mathbf{r} - 
    \omega_{p}t)}{\boldsymbol \epsilon}_{p}+c.c.,
\end{eqnarray*}
where $\mathcal{E}_{0}$, $\mathcal{E}_{p}$, and $\mathcal{E}_{c}$ are the
field amplitudes for the pump, the probe, and the conjugate, respectively, and ${\boldsymbol \epsilon}_{i}$ are unit vectors describing the polarization of
the fields.  We assume that the pump couples the transitions $| 1
\rangle\rightarrow | 3 \rangle$ and $| 2\rangle\rightarrow | 4 \rangle$, the
probe couples transition $| 2 \rangle\rightarrow | 3 \rangle$, the conjugate
couples transition $| 1 \rangle\rightarrow | 4 \rangle$, and that the
transitions $| 1 \rangle\rightarrow | 2 \rangle$ and $| 3 \rangle\rightarrow |
4 \rangle$ are not dipole allowed. With these assumptions and using the
rotating-wave approximation, the equations of motion for the density matrix
elements take the form
\begin{widetext}
\begin{eqnarray}
    \dot{\sigma}_{11}&=&\frac{i}{2}(\Omega^{*}_{1}e^{-i\mathbf{k}_0 \cdot
    \mathbf{r}}\sigma_{31}+
    \Omega^{*}_{p}e^{-i\mathbf{k}_p \cdot
    \mathbf{r}}\sigma_{41}-\Omega_{1}e^{i\mathbf{k}_0 \cdot
    \mathbf{r}}\sigma_{13}-\Omega_{p}e^{i\mathbf{k}_p \cdot \mathbf{r}}\sigma_{14})%\nonumber\\
    %&& 
    +\Gamma_{13}\sigma_{33}+\Gamma_{14}\sigma_{44}\label{dbllambdaeq1}\\
    \dot{\sigma}_{22}&=&\frac{i}{2}(\Omega^{*}_{c}e^{-i\mathbf{k}_c \cdot
    \mathbf{r}}\sigma_{32}+
    \Omega^{*}_{2}e^{-i\mathbf{k}_0 \cdot
    \mathbf{r}}\sigma_{42}-\Omega_{c}e^{i\mathbf{k}_c \cdot
    \mathbf{r}}\sigma_{23}-\Omega_{2}e^{i\mathbf{k}_0 \cdot \mathbf{r}}\sigma_{24})%\nonumber\\
    %&&
    +\Gamma_{23}\sigma_{33}+\Gamma_{24}\sigma_{44}\\
    \dot{\sigma}_{33}&=&\frac{i}{2}(\Omega_{1}e^{i\mathbf{k}_0 \cdot
    \mathbf{r}}\sigma_{13}+\Omega_{c}e^{i\mathbf{k}_c \cdot \mathbf{r}}\sigma_{23}-
    \Omega^{*}_{1}e^{-i\mathbf{k}_0 \cdot
    \mathbf{r}}\sigma_{31}-\Omega^{*}_{c}e^{-i\mathbf{k}_c \cdot \mathbf{r}}\sigma_{32})%\nonumber\\
    %&&
    -\Gamma_{3}\sigma_{33}\\
    \dot{\sigma}_{44}&=&\frac{i}{2}(\Omega_{p}e^{i\mathbf{k}_p \cdot
    \mathbf{r}}\sigma_{14}+\Omega_{2}e^{i\mathbf{k}_0 \cdot \mathbf{r}}\sigma_{24}-
    \Omega^{*}_{p}e^{-i\mathbf{k}_p \cdot
    \mathbf{r}}\sigma_{41}-\Omega^{*}_{2}e^{-i\mathbf{k}_0 \cdot \mathbf{r}}\sigma_{42})%\nonumber\\
    %&&
    -\Gamma_{4}\sigma_{44}\\
    \dot{\sigma}_{43}&=&\frac{i}{2}(\Omega_{2}e^{i\mathbf{k}_0 \cdot
    \mathbf{r}}\sigma_{23}+
    \Omega_{p}e^{i\mathbf{k}_p \cdot
    \mathbf{r}}\sigma_{13}-\Omega^{*}_{1}e^{-i\mathbf{k}_0 \cdot \mathbf{r}}\sigma_{41}-
    \Omega^{*}_{c}e^{-i\mathbf{k}_c \cdot \mathbf{r}}\sigma_{42})%\nonumber\\
    %&&
    +(i\Delta_{2}-i\Delta_{1}-\gamma_{43})\sigma_{43}\\
    \dot{\sigma}_{42}&=&\frac{i}{2}(\Omega_{2}e^{i\mathbf{k}_0 \cdot
    \mathbf{r}}\sigma_{22}+
    \Omega_{p}e^{i\mathbf{k}_p \cdot
    \mathbf{r}}\sigma_{12}-\Omega_{c}e^{i\mathbf{k}_c \cdot
    \mathbf{r}}\sigma_{43}-\Omega_{2}e^{i\mathbf{k}_0 \cdot \mathbf{r}}\sigma_{44})%\nonumber\\
    %&&
    +(i\Delta_{2}-i\delta-\gamma_{42})\sigma_{42}\\
    \dot{\sigma}_{41}&=&\frac{i}{2}(\Omega_{2}e^{i\mathbf{k}_0 \cdot
    \mathbf{r}}\sigma_{21}+\Omega_{p}e^{i\mathbf{k}_p \cdot \mathbf{r}}\sigma_{11}-
    \Omega_{1}e^{i\mathbf{k}_0 \cdot
    \mathbf{r}}\sigma_{43}-\Omega_{p}e^{i\mathbf{k}_p \cdot \mathbf{r}}\sigma_{44})%\nonumber\\
    %&&
    +(i\Delta_{2}-\gamma_{43})\sigma_{41}\\
    \dot{\sigma}_{32}&=&\frac{i}{2}(\Omega_{c}e^{i\mathbf{k}_c \cdot
    \mathbf{r}}\sigma_{22}+\Omega_{1}e^{i\mathbf{k}_0 \cdot \mathbf{r}}\sigma_{12}-
    \Omega_{c}e^{i\mathbf{k}_c \cdot
    \mathbf{r}}\sigma_{33}-\Omega_{2}e^{i\mathbf{k}_0 \cdot \mathbf{r}}\sigma_{34})%\nonumber\\
    %&&
    +(i\Delta_{1}-i\delta-\gamma_{32})\sigma_{32}\\
    \dot{\sigma}_{31}&=&\frac{i}{2}(\Omega_{c}e^{i\mathbf{k}_c \cdot
    \mathbf{r}}\sigma_{21}+\Omega_{1}e^{i\mathbf{k}_0 \cdot \mathbf{r}}\sigma_{11}-
    \Omega_{1}e^{i\mathbf{k}_0 \cdot
    \mathbf{r}}\sigma_{33}-\Omega_{p}e^{i\mathbf{k}_p \cdot \mathbf{r}}\sigma_{34})%\nonumber\\
    %&&
    +(i\Delta_{1}-\gamma_{31})\sigma_{31}\\
    \dot{\sigma}_{21}&=&\frac{i}{2}(\Omega^{*}_{c}e^{-i\mathbf{k}_c \cdot
    \mathbf{r}}\sigma_{31}+\Omega^{*}_{2}e^{-i\mathbf{k}_0 \cdot \mathbf{r}}\sigma_{41}-
    \Omega_{1}e^{i\mathbf{k}_0 \cdot
    \mathbf{r}}\sigma_{23}-\Omega_{p}e^{i\mathbf{k}_p \cdot \mathbf{r}}\sigma_{24})%\nonumber\\
    %&&
    +(i\delta-\gamma_{21})\sigma_{21}\label{dbllambdaeq10},
\end{eqnarray}
\end{widetext}
where $\Omega_{1}$ and $\Omega_{2}$ give the Rabi frequencies for the two
transitions coupled by the single pump, $\Omega_{p}$ the Rabi frequency for
the transition coupled by the probe, and $\Omega_{c}$ the Rabi frequency for
the transition coupled by the conjugate. In order to completely eliminate the
explicit time dependence when doing the rotating-wave approximation, the
frequencies of the pump, probe, and conjugate fields need to satisfy the
relation
\begin{equation*}
    2\omega_{0}=\omega_{p}+\omega_{c},
\end{equation*}
which is just an energy conservation condition for the 4WM process.

In order to solve the above equations and obtain analytical expressions, we
assume that the probe and conjugate fields are weak fields, such that we only
keep terms to first order in $\Omega_{p}$ and $\Omega_{c}$.  In this case, the
polarization of the medium at frequency $\omega_{i}$ (where $i$ indicates
probe or conjugate frequency) can be divided into two different terms, one
that is proportional to the field at frequency $\omega_{i}$ and one that is
proportional the field at frequency $2\omega_{0}-\omega_{i}$, such that
%We call the susceptibilities for these two terms the direct term and the
%cross term, respectively.  The polarization of the medium then takes the form
%
\begin{eqnarray} P(\omega_{p}) & = &
    \epsilon_{0}\chi_{pp}(\omega_{p})\mathcal{E}_{p}e^{i \mathbf{k}_p \cdot
    \mathbf{r}} \nonumber\\
    &&+\epsilon_{0}\chi_{pc}(\omega_{p})\mathcal{E}_{c}^{*}e^{i(2\mathbf{k}_0
    - \mathbf{k}_c) \cdot \mathbf{r}} \label{probepol}\\ P(\omega_{c}) & = &
    \epsilon_{0}\chi_{cc}(\omega_{c})\mathcal{E}_{c}e^{i\mathbf{k}_c \cdot
    \mathbf{r}} \nonumber\\ &&+
    \epsilon_{0}\chi_{cp}(\omega_{c})\mathcal{E}_{p}^{*}e^{i(2\mathbf{k}_0 -
    \mathbf{k}_p) \cdot \mathbf{r}}.  \label{conjpol} \end{eqnarray}
In doing this, we have introduced the susceptibility of the atomic medium
$\chi_{ij}$, which completely characterizes the response of the atomic system
for a given field. The direct susceptibilities $\chi_{pp,cc}$ act as the
effective linear susceptibilities for the probe and conjugate, respectively;
the cross-susceptibilities $\chi_{pc,cp}$ are responsible for the 4WM process.

In addition to the approximations mentioned above, we assume the
dipole moments for the two pump transitions to be equal, such that
$\Omega_{1}=\Omega_{2}\equiv\Omega$.  Under these approximations we can solve
the density matrix equations, Eqs.~(\ref{dbllambdaeq1}-\ref{dbllambdaeq10}), to all orders in the pump field ($\Omega$) and in steady-state condition for $\sigma_{41}$ and
$\sigma_{32}$ to find that
\begin{eqnarray}
    \chi_{pp}&=&\frac{i\mathcal{N}|d_{23}|^{2}\xi_{41}^{*}}{\epsilon_{0}\hbar
    D^{*}}
    \left[\frac{\xi_{21}^{*}}{\xi_{42}^{*}}\sigma_{22,44}+\frac{\xi_{43}^{*}}{\xi_{31}^{*}}\sigma_{11,33}\right.\nonumber\\*
    &&-\left.\left(\frac{\xi_{21}^{*}+\xi_{43}^{*}}{\xi_{41}^{*}}+\frac{\xi_{21}^{*}\xi_{43}^{*}}{|\Omega|^{2}/4}\right)\sigma_{22,33}\right]\label{dbllambdachi1}\\
    \chi_{cc}&=&\frac{i\mathcal{N}|d_{14}|^{2}\xi_{32}^{*}}{\epsilon_{0}\hbar
    D}
    \left[\frac{\xi_{43}}{\xi_{42}^{*}}\sigma_{22,44}+\frac{\xi_{21}}{\xi_{31}^{*}}\sigma_{11,33}\right.\nonumber\\*
    &&-\left.\left(\frac{\xi_{21}+\xi_{43}}{\xi_{32}^{*}}+\frac{\xi_{21}\xi_{43}}{|\Omega|^{2}/4}\right)\sigma_{11,44}\right]\label{dbllambdachi2}\\
    \chi_{pc}&=&\frac{i\mathcal{N}d_{14}d_{23}\xi_{41}^{*}\Omega^{2}}{\epsilon_{0}\hbar
    D^{*}|\Omega|^{2}}
    \left[\frac{\xi_{21}^{*}}{\xi_{31}}\sigma_{11,33}+\frac{\xi_{43}^{*}}{\xi_{42}}\sigma_{22,44}\right.\nonumber\\*
    &&+\left.\left(\frac{\xi_{21}^{*}+\xi_{43}^{*}}{\xi_{41}^{*}}\right)\sigma_{11,44}\right]\label{dbllambdachi3}\\
    \chi_{cp}&=&\frac{i\mathcal{N}d_{14}d_{23}\xi_{32}^{*}\Omega^{2}}{\epsilon_{0}\hbar
    D|\Omega|^{2}}
    \left[\frac{\xi_{43}}{\xi_{31}}\sigma_{11,33}+\frac{\xi_{21}}{\xi_{42}}\sigma_{22,44}\right.\nonumber\\*
    &&+\left.\left(\frac{\xi_{21}+\xi_{43}}{\xi_{32}^{*}}\right)\sigma_{22,33}\right]\label{dbllambdachi4},
\end{eqnarray}
where we have defined
\begin{equation*}
    D=(\xi_{43}+\xi_{21})(\xi_{32}^{*}+\xi_{41})+\frac{\xi_{32}^{*}\xi_{41}\xi_{43}\xi_{21}}{|\Omega|^{2}/4},
\end{equation*}
the population differences
\begin{eqnarray}
    \sigma_{11,33}\equiv\sigma_{11}-\sigma_{33}=\frac{|\xi_{31}|^{2}}{|\Omega|^{2}+|\xi_{31}|^{2}+|\xi_{42}|^{2}}\label{pop1}\\
    \sigma_{11,44}\equiv\sigma_{11}-\sigma_{44}=\frac{|\xi_{31}|^{2}}{|\Omega|^{2}+|\xi_{31}|^{2}+|\xi_{42}|^{2}}\label{pop2}\\
    \sigma_{22,33}\equiv\sigma_{22}-\sigma_{33}=\frac{|\xi_{42}|^{2}}{|\Omega|^{2}+|\xi_{31}|^{2}+|\xi_{42}|^{2}}\label{pop3}\\
    \sigma_{22,44}\equiv\sigma_{22}-\sigma_{44}=\frac{|\xi_{42}|^{2}}{|\Omega|^{2}+|\xi_{31}|^{2}+|\xi_{42}|^{2}},\label{pop4}
\end{eqnarray}
and the complex decay rates
\begin{eqnarray*} 
    \xi_{43}& =& i(\Delta_{2}-\Delta_{1})-\gamma\\
    \xi_{42}& =& i(\Delta_{2}-\delta)-\frac{\gamma}{2}\\
    \xi_{41}& =& i\Delta_{2}-\frac{\gamma}{2}\\
    \xi_{32}& =& i(\Delta_{1}-\delta)-\frac{\gamma}{2}\\
    \xi_{31}& =& i\Delta_{1}-\frac{\gamma}{2}\\
    \xi_{21}& =& i\delta-\gamma_{c}.
\end{eqnarray*}

In deriving these equations, we have assumed that the total decay rate out of
the excited states are the same and with equal branching ratios to the two
ground states; that is, $\Gamma_{4}=\Gamma_{3}\equiv\gamma$ and
$\Gamma_{14}=\Gamma_{24}=\Gamma_{13}=\Gamma_{14}=\gamma/2$.  In addition, we
assume that the additional dipole dephasing term $\gamma_{ij}^{c}$ is only
significant for the ground state coherence, such that
$\gamma_{ij\neq12}^{c}=0$ and $\gamma_{12}^{c}\equiv\gamma_{c}$.

\subsection{Index of refraction for the pump} 

In the above derivation of the susceptibilities for the
probe and the conjugate, we have implicitly set $\mathcal{E}_0$, the amplitude
for the pump electric field, to be a constant throughout the medium. This
assumes that the pump is neither dephased or absorbed. In practice, the pump
detuning is large enough to neglect absorption, but even a small index of
refraction may have a substantial impact on the phase-matching condition of
the 4WM process. The effect of refraction on the pump is to multiply
$\mathcal{E}_0$ (and therefore $\Omega$) by a running phase factor as the pump
propagates. This can be taken into account by simply replacing $\mathbf{k}_0$
in $\Delta \mathbf{k}$ by $n_0 \mathbf{k}_0$, where $n_0$ is the index of
refraction. The index of refraction created by the population $\mathcal{N}_i$
of the ground state $i$ can be estimated to be $n_0 = \sqrt{1 + \chi} \simeq 1
+ \chi/2$ with
$$ \chi = -\frac{\mathcal{N}_i d^2}{\varepsilon_0 \hbar}
\frac{\Delta_i}{\Delta_i^2 + \gamma^2/4}, $$
where $d$ is the dipole matrix element of the transition for large detunings and
$\Delta_i$ is the detuning, which is taken to be much larger than the hyperfine
splitting of the excited state.

\bibliography{phasematch.bib}

\end{document}